\documentclass[letterpaper,12pt]{article}
\usepackage[margin=1in]{geometry}

\usepackage{mathpazo}

\usepackage[comma]{natbib}
\setcitestyle{aysep={}}

\usepackage{amsmath}
\usepackage{titlesec}
\titleformat*{\section}{\filcenter\MakeUppercase}
\titleformat*{\subsection}{\filcenter}
\usepackage{graphicx}
\usepackage{caption}

\DeclareMathOperator{\logit}{logit}

\newcommand*{\sss}{\scriptscriptstyle}

\DeclareSymbolFont{yhlargesymbols}{OMX}{yhex}{m}{n}
\DeclareMathAccent{\wideparen}{\mathord}{yhlargesymbols}{"F3}

\usepackage{abstract}

\makeatletter
\renewcommand*{\l@figure}[2]{%
\setlength\@tempdima{2.3em}%
\noindent\hspace*{1.5em}#1\hfil\newline }
\makeatother

\usepackage{authblk}

\title{A test for monitoring under- and overtreatment in Dutch hospitals}
\author{Oliver Urs Lenz \& Daniel L Oberski}
\date{}

\begin{document}

\maketitle
    \begin{onecolabstract}      \begin{minipage}{1.0\linewidth}

Over- and undertreatment harm patients and society and confound other healthcare quality measures. Despite a growing body of research covering specific conditions, we lack tools to systematically detect and measure over- and undertreatment in hospitals. We demonstrate a test used to monitor over- and undertreatment in Dutch hospitals, and illustrate its results applied to the aggregated administrative treatment data of 1,836,349 patients at 89 hospitals in 2013. We employ a random effects model to create risk-adjusted funnel plots that account for natural variation among hospitals, allowing us to estimate a measure of overtreatment and undertreatment when hospitals fall outside the control limits. The results of this test are not definitive, findings were discussed with hospitals to improve the model and to enable the hospitals to make informed treatment decisions.

      \end{minipage}

\end{onecolabstract}

\section{Introduction}

Undertreatment and overtreatment are well-documented problems of health care quality. While the negative consequences of undertreatment for patients are obvious, overtreatment can also harm patients \citep{franks1992gatekeeping,bangma2007overdiagnosis,esserman2013overdiagnosis}. Moreover, overtreatment increases the costs of health care to society at large. By one estimate, between 6\% and 8\% of \emph{total} spending on health care in the United States in 2011 was due to overtreatment, amounting to between \$18 and \$44 billion \citep{berwick2012eliminating}. Eliminating these costs is generally a preferable alternative to other types of cutbacks such as lowering insurance coverage.

Apart from being harmful in their own right, over- and undertreatment also affect subsequent quality evaluation measures of specific negative outcomes. In-hospital mortality, readmission, and long-stay rates all depend on the prior question of how easily a person becomes a patient at all. For example, if a hospital diagnoses additional patients that would otherwise not have been treated, these `light' patients will lower its mortality, readmission, and long-stay rates. Conversely, if a hospital, for whatever reason, is more disinclined to treat people with a particular diagnosis, it may be the more severe cases that remain. Such differences will impact quality evaluations that compare health care providers (such as \citealt{coory2008using,bardsley2009using,jarman2010hospital}).

Clearly we want to be able to detect and measure under- and overtreatment. Several studies have examined overtreatment of particular diagnoses, especially prostate cancer (see \citealt{bratt2006watching,bangma2007overdiagnosis,heijnsdijk2009overdetection} and references therein), but also thyroid cancer \citep{lee2012incidence}, cancer in general \citep{esserman2013overdiagnosis}, hypertension \citep{furberg1994overtreatment}, asthma \citep{caudri2011asthma}, diabetes mellitus \citep{sutin2010diabetes}, and even childbirth \citep{albers2005overtreatment}. But to our knowledge, we lack more general approaches of measuring over- and undertreatment.

This article demonstrates a test that is used to monitor both over- and undertreatment across diagnoses. Recognizing that natural variation between providers can cause significant differences, we use funnel plots based on random effects models \citep{spiegelhalter05,ohlssen07} of casemix-adjusted incidence rates. We apply these funnel plots to administrative data from Dutch health care providers and demonstrate how they may be used to follow up on unusual incidence rates. While funnel plots are well suited for this purpose \citep{vandishoeck2011displaying}, they still require careful interpretation \citep{neuburger2011funnel,seaton2013what,mohammed2013statistical,shahian2015what}. The test does not provide definitive evidence of over- or undertreatment, but is used as an indicator, the results of which are discussed with the hospitals themselves.

\section{Methods}

\subsection{Data}

The administrative data discussed in this article comprises all insurance claims made by 89 health care providers with one major health insurer in the Netherlands for the year 2013, involving 1,836,349 patients. Aggregate statistics on these claims were provided by i2i --- Intelligence to Integrity, a company that produces health care quality reports for Dutch hospitals. Included were all providers categorized as regional, general, top-clinical or academic. This excludes categorical hospitals and independent treatment centres, which treat a limited set of conditions. The claims are categorized by diagnosis -- 2357 in total, divided over 29 medical specialities. Conforming to privacy regulations, only aggregate statistics were obtained by the authors.

\subsection{Funnel plots}

Across hospitals, there is variation in the treatment of medical conditions. Examples of treatment choices include whether to admit into hospital, whether to operate, and when to discharge a patient. Much of this variation will exist for clinical reasons -- that is, it will occur ``naturally'' due to the distribution of patients and medical professionals. However, there may also be exceptional deviations that cannot be explained clinically. An early warning system for such exceptional deviations must therefore account for both natural variation and the possibility of over- and undertreatment. We build such a system for the Netherlands using funnel plots.

The funnel plot \citep{spiegelhalter05} is a type of control chart in which the control limits account for both statistical and natural variation among the providers \citep{ohlssen07}. Control charts such as the funnel plot have been shown to lead to better decision-making regarding the follow-up of unusual performance outcomes \citep{marshall2004randomized}.

Our funnel plots are constructed on the incidence rates per diagnosis. When a provider appears above or below the control limits it can be said to treat patients more or less readily than other providers. However, this need not be absolute over- or undertreatment. Firstly, the surplus or deficit may be relative, due to e.g. providers having different registration practices, attributing different diagnoses to similar patients. Secondly, we may not want to call a surplus or deficit over- or undertreatment if it has an acceptable explanation. In either case, it is still desirable to be able to identify the relevant providers. However, it is clearly important to distinguish these cases from genuine over- and undertreatment. This requires evaluation of the reason for a surplus or deficit by the hospital itself. The Discussion further elaborates these points.

\subsection{Setting  control limits with random effects models}

Per diagnosis and per provider $i \in I$, we define incidence as the number of patients $o_i$ divided by the total number of potential patients $n_i$, and compare this with the expected number of patients $e_i$ based on casemix adjustment. Given these figures, funnel plots with limits determined by random effects models provide us with a technique to determine whether the incidence of a provider deviates from the range of `common cause' variability \citep{spiegelhalter05,ohlssen07}.

Following \cite[pp. 868--70]{ohlssen07}, we model the excess log-odds ratio of becoming a patient,
\begin{equation}\label{eq:ylogit}
	y_i := \logit(o_i / n_i) - \logit(e_i / n_i).
\end{equation}
This excess log-odds ratio is modelled using a normal approximation,
\begin{equation}
	y_i \sim \mathcal{N}(\mu_i, s_i),
\end{equation}
where its standard deviation is taken to be estimated by $s_i = \sqrt{1/o_i + 1/(n_i-o_i)}$. After the implicit casemix adjustment in Eq.~\ref{eq:ylogit}, true hospital excesses $\gamma_i$ may still vary due to a number of small random factors, leading to a normal random effects model \citep{marshall2004statistical},
\begin{equation}
	\mu_i \sim \mathcal{N}(\mu, \tau). 
\end{equation}

For ease of computation from the very large database of claims, the parameters $\mu$ and $\tau$ are estimated from the observed data $y_i$ using the  method-of-moments estimator \citep{dersimonian1986meta}. We then create a funnel plot \citep{spiegelhalter05} by plotting estimates of the true excess log-odds ratios,
\begin{equation}
	\hat{\mu}_i = w^*_i\sum_{\sss i \in I}{y_i}/\sum_{\sss i \in I}{w^*_i},
\end{equation}
where $w^*_i := (s_i^2 + \hat{\tau}^2)^{-1}$, against the precision $s_i^{-1}$ with funnel lines at
\begin{equation}
	\hat{\mu} \pm c(s_i^2 + \hat{\tau}^2).
\end{equation}
where $c$ is the number of standard deviations that one wants to consider. We choose $c=2$ (2 sigma), corresponding to a $95\%$ prediction limit.

While we have access to the number $o_i$ of patients treated per provider $i$ and per diagnosis, there is no immediate way of knowing the number of non-patients, and, by extension, the total number of potential patients $n_i$. We estimate $n_i$ by distributing the total number of people over those providers that have treated the diagnosis, according to ``market share''. This is determined on the basis of the total number of patients of the speciality, and is adjusted for sex, age group, and socio-economic status.

\section{Results}

The calculations as detailed in the previous section were run for all diagnoses. We illustrate the results with six diagnoses that have been highlighted in the literature for their variation \citep{wennberg80,schooten08}. The first three diagnoses exemplify less severe conditions that under some circumstances may not be serious enough to warrant hospital treatment: (i) haemorrhoids, (ii) acute otitis media (AOM), otitis media with effusion (OME), or Eustachian tube dysfunction, and (iii) nasal septum deviations. The remaining three conditions are serious, especially when left untreated. At the same time, the corresponding treatments may be high-risk and may not improve quality of life for every patient. These are (iv) cataract, (v) prostate carcinoma, and (vi) cholecystisis/cholelithiasis.

\begin{figure}

\vspace{-3\baselineskip}
\includegraphics[height=.35\textheight]{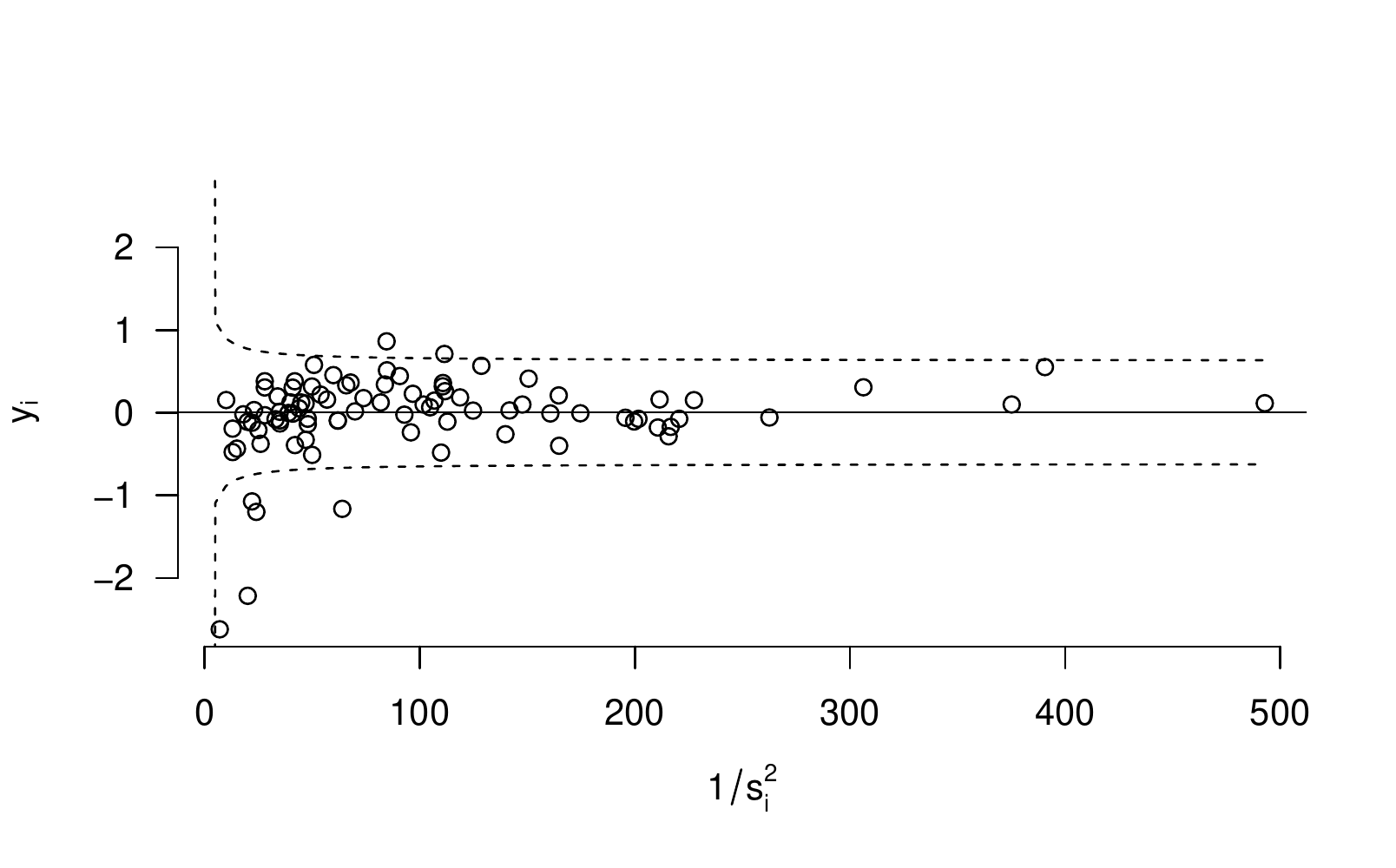}

\vspace{-3\baselineskip}
\includegraphics[height=.35\textheight]{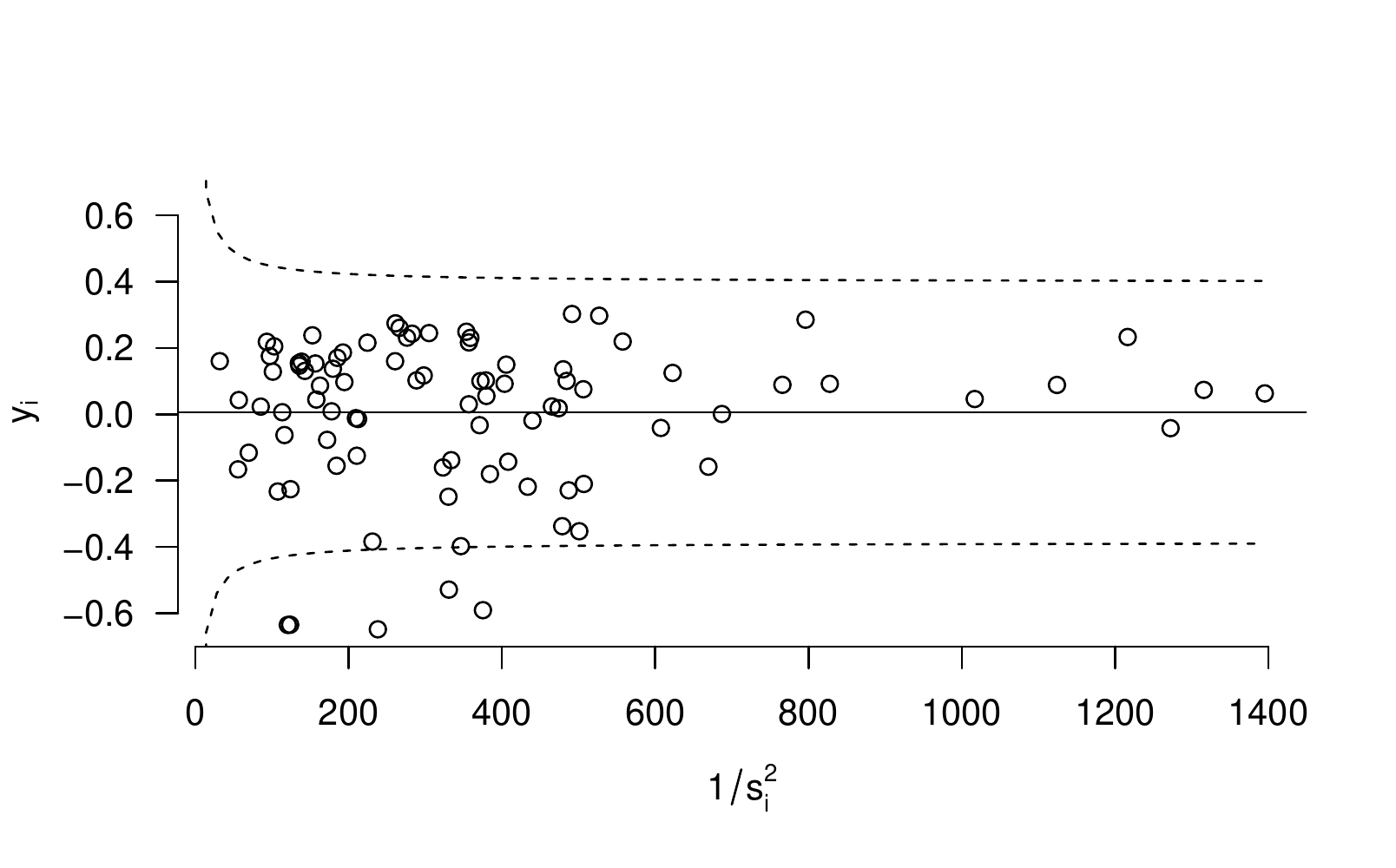}

\vspace{-3\baselineskip}
\includegraphics[height=.35\textheight]{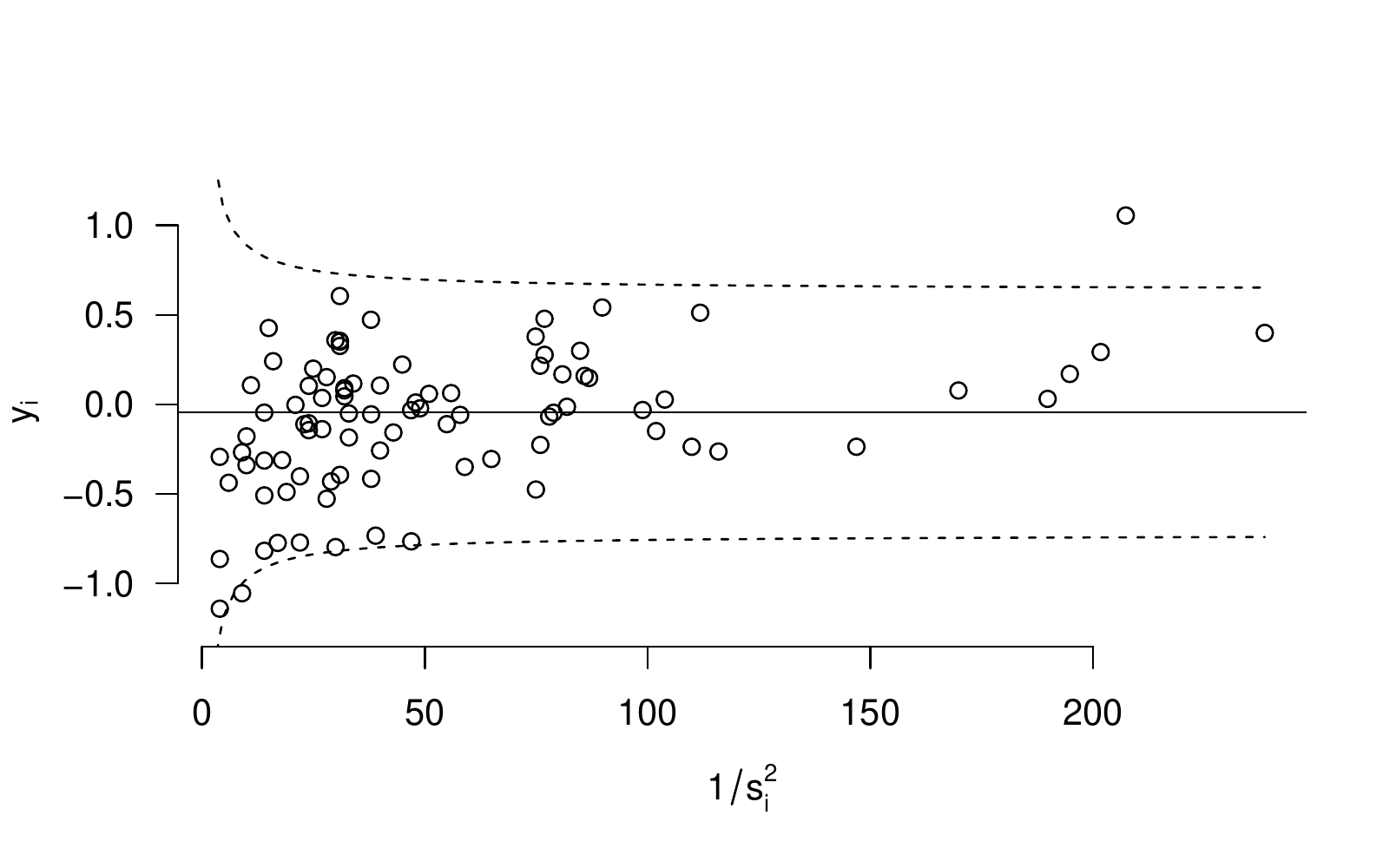}

\caption{Funnel plots for three relatively light conditions. Top: haemorrhoids, middle: AOM, OME, and Eustachian tube dysfunction, bottom: nasal septum deviations.}
\label{fig:funnel-light}
\end{figure}

\begin{figure}

\vspace{-3\baselineskip}
\includegraphics[height=.35\textheight]{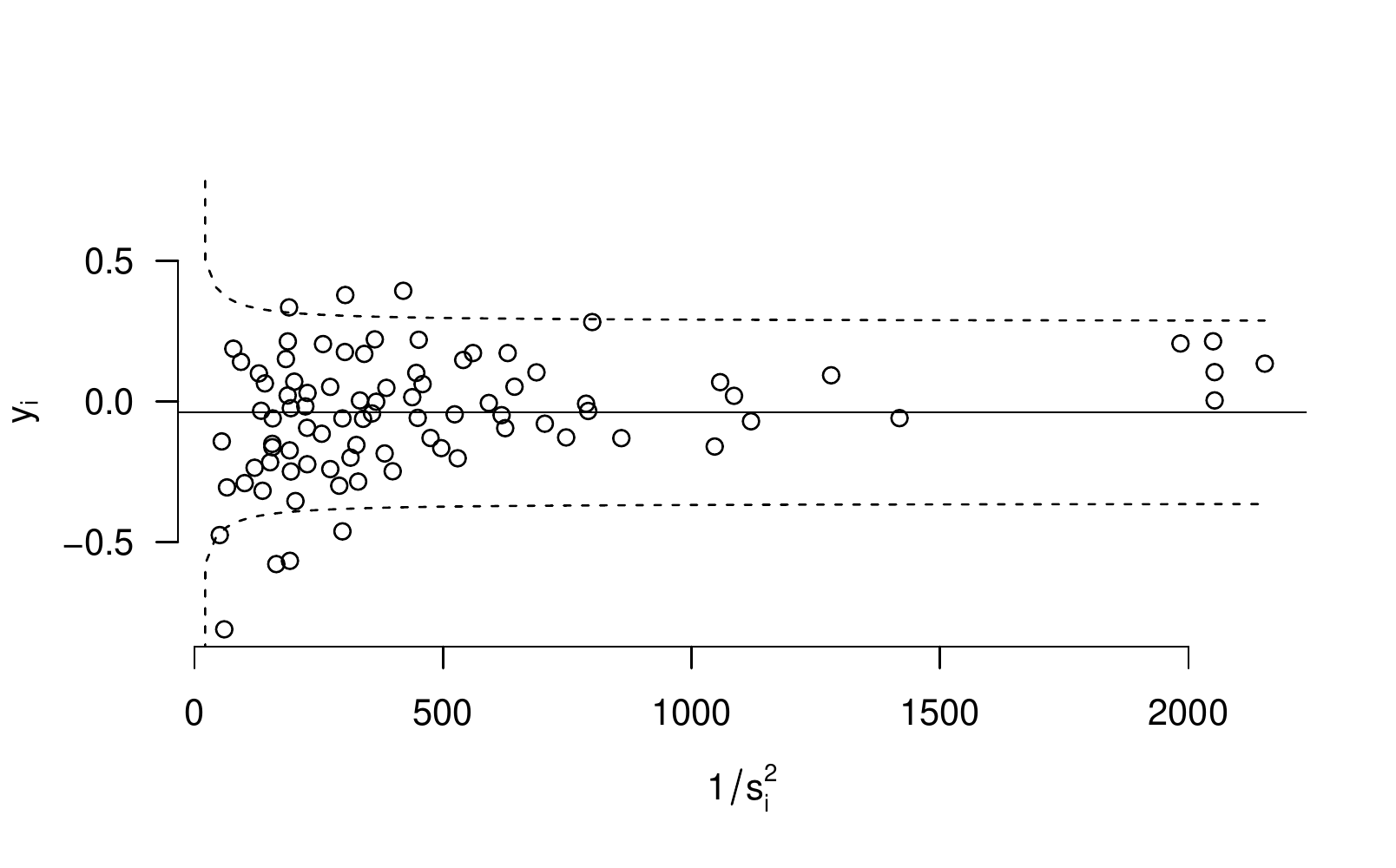}

\vspace{-3\baselineskip}
\includegraphics[height=.35\textheight]{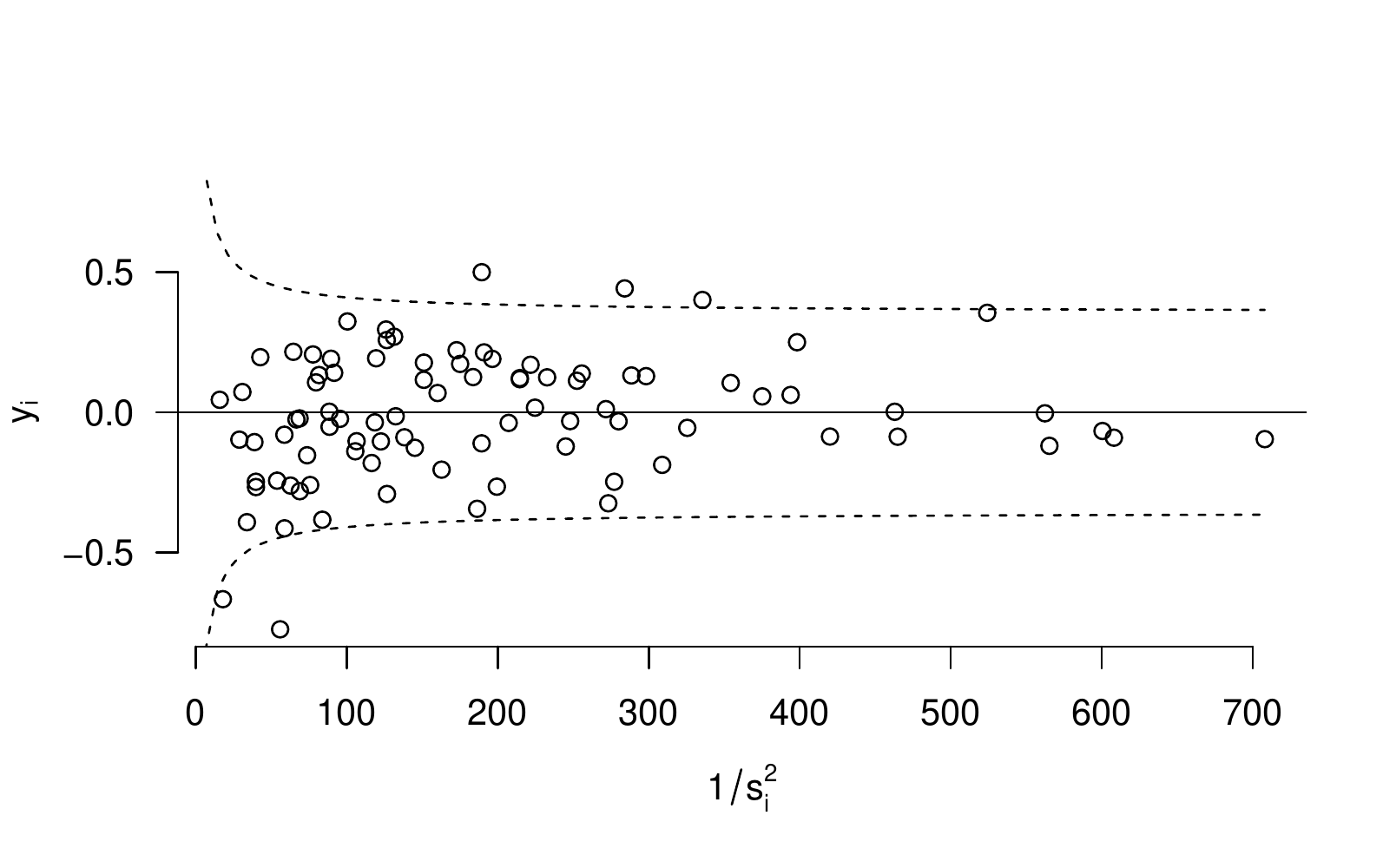}

\vspace{-3\baselineskip}
\includegraphics[height=.35\textheight]{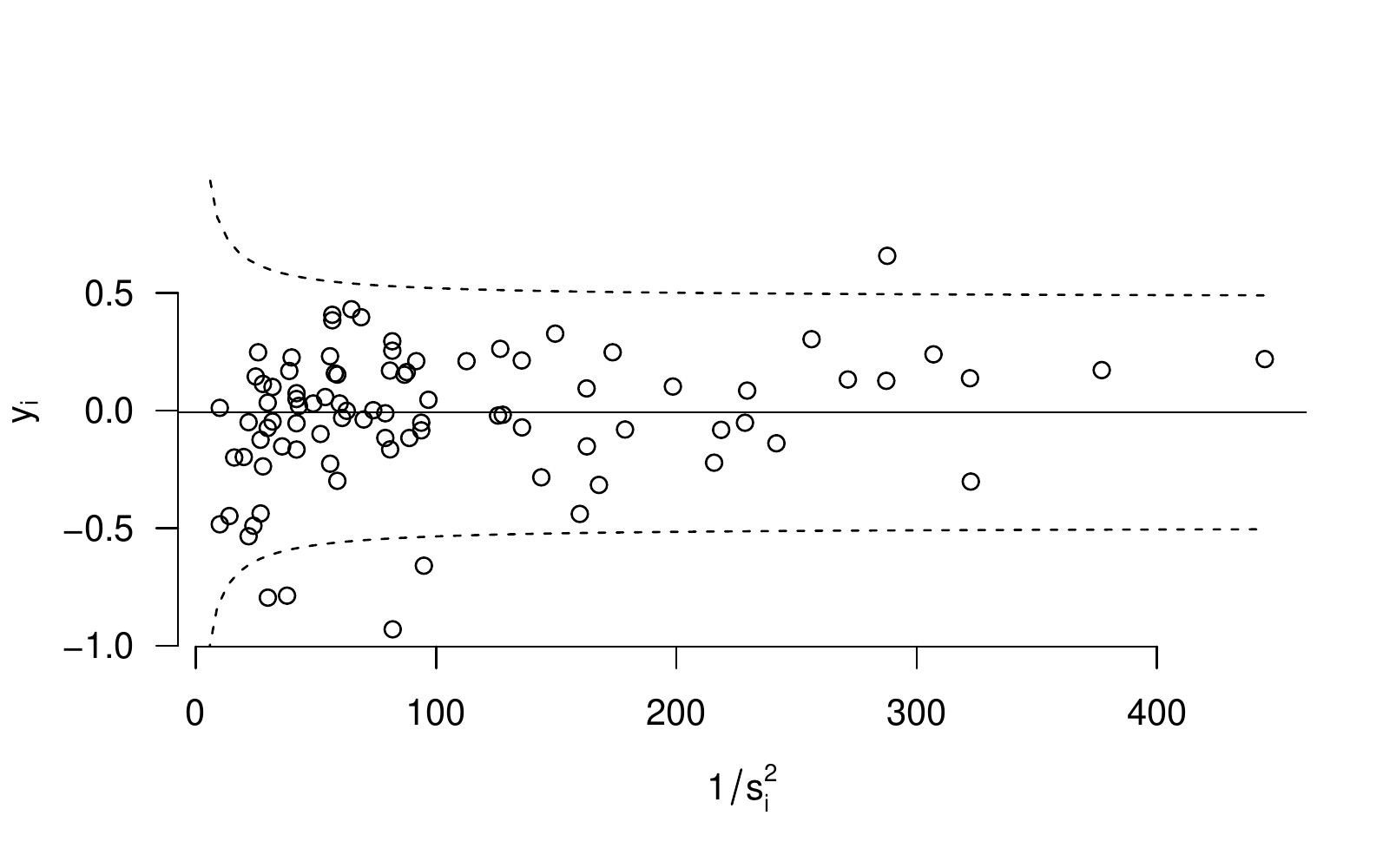}

\caption{Funnel plots for three serious conditions. Top: cataract, middle: prostate carcinoma, bottom: cholecystisis/cholelithiasis.}
\label{fig:funnel-severe}
\end{figure}

Figure~\ref{fig:funnel-light} displays the resulting funnel plots for the less severe conditions. For all three plots, most hospitals are within the control limits. However, for haemorrhoids and nasal septum deviations, there are respectively two and one providers with unusually many patients, even after adjusting for casemix and natural variation. For haemorrhoids and AOM, OME, and Eustachian tube dysfunction, a number of hospitals have unusually few patients.

The funnel plots in Figure~\ref{fig:funnel-severe} show similar patterns. For these conditions, there are respectively four, two, and four hospitals with unusually few patients. Considering the gravity of these conditions, follow-up may be needed. Towards the upper end of the plots, we find respectively three, three and one providers with unusually many patients, which may be justified, but could also indicate overtreatment. This, too, warrants investigation by the relevant hospitals.

\section{Discussion}

The results show that while hospitals vary with respect to patient numbers, in most cases the variation is natural. Nonetheless, for each of the conditions examined, some providers fall outside the control limits. Funnel plots constitute a useful tool for identifying hospitals with higher or lower patient numbers requiring explanation. Scrutinizing these cases is a first step towards reducing the costs and improving the quality of health care.

A strength of our method is that since it is based on administrative data, it can be automated to apply across hospitals and diagnoses. Moreover, it is an objective test to determine whether a hospital is really performing worse than other hospitals. As such, it provides an alternative to simple ranking indicators, which may be misleading in as much as the relative ranking is not statistically significant \citep{leleu2014hospital}.

While it is possible to obtain initial results with funnel plots relatively easily, they should not be used in isolation, because some providers may fall outside the control limits for excellent reasons. Moreover, recent studies have emphasized the limitations of relying solely on funnel plots to compare hospitals \citep{neuburger2011funnel,seaton2013what,mohammed2013statistical,shahian2015what}. Outcome assessment using funnel plots should therefore take place within a wider framework that gives agency to health care providers. The results produced here were embedded into such a system: feedback was given to participating providers allowing them to employ their own expertise to make informed decisions. This was done as part of an ongoing process where regular updates on the basis of new treatment figures enable the evaluation of these decisions.

Interviews with doctors and hospital staff revealed a number of additional factors that were fed back into the analysis as additional casemix adjustments. Closer inspection showed that certain apparent cases of absolute over- or undertreatment could be explained as dispersion among hospitals or diagnoses. In a number of instances, similar patients were attributed different diagnoses by different providers. In order to remove this effect, the relevant diagnoses were combined. Another potential complication is that a provider may be seen to undertreat if patients with certain conditions are drawn to a specialized clinic that happens to be nearby, or conversely, to overtreat, if it itself provides specific expertise. In some cases, it was possible to compensate for this by re-attributing patients. A more radical solution would be to combine certain providers for specific conditions in order to determine collective under- or overtreatment.

There are three main limitations to our study. First, we analysed data from only one health insurer. While encompassing a large number of patients, it may be the case that we missed some systematic variation across hospitals due to this selection. Future studies might analyze data from all insurers or from a random sample of claims. Second, economical accounts such as \cite[][Section 5]{mcguire2000physician} and \citealt{alger2006theory} identify mechanisms that might engender general overtreatment by doctors. The argument that there is a general tendency for overtreatment has also been made for specific conditions such as thyroid cancer \citep{lee2012incidence}, for example due to the liberal application of diagnostic tools. Such general over- (or under-)treatment occurring across all hospitals is not detected by our method. Third, our approach is blind to under- and overtreatment on an individual level, as for instance identified for asthma \citep{caudri2011asthma}, if this cancels out at the hospital-level.

\section{Acknowledgments}

We are grateful to Luca Schippa for assisting with the data preparation and to Michel Taal for comments on the manuscript.

\bibliographystyle{oliver}
\bibliography{funnel_plots}

\end{document}